\documentclass[useAMS,usenatbib]{mn2e} 
\usepackage{graphicx}

\def \src {MXB\, 1658--298}

\title[X-ray eclipse timing of \src]{Indication of a massive circumbinary planet orbiting the Low Mass X-ray Binary \src}

\author[Jain et al.]{Chetana Jain$^{1}$\thanks{E-mail: chetanajain11@gmail.com (CJ)}, Biswajit Paul$^{2}$, Rahul Sharma$^{3}$, Abdul Jaleel$^{3}$ and Anjan Dutta$^{3}$\\
$^{1}$Hans Raj College, University of Delhi, Delhi 110007, India\\
$^{2}$Raman Research Institute, Sadashivnagar, C. V. Raman Avenue, Bangalore 560080, India\\
$^{3}$Department of Physics and Astrophysics, University of Delhi, Delhi 110007, India}

\begin{document}

\date{ }

\pagerange{\pageref{firstpage}--\pageref{lastpage}} \pubyear{ }

\maketitle

\label{firstpage}

\begin{abstract}

We present an X-ray timing analysis of the transient X-ray binary \src, using data obtained from the $RXTE$ and $XMM-Newton$ observatories. We have made 27 new mid eclipse time measurements from observations made during the two outbursts of the source. These new measurements have been combined with the previously known values to study long term changes in  orbital period of the binary system. We have found that the mid-eclipse timing record of \src~ is quite unusual. The long term evolution of mid-eclipse times indicates an overall orbital period decay with a time scale of -- 6.5(7) $\times$ 10$^{7}$ year. Over and above this orbital period decay, the O-C residual curve also shows a periodic residual on shorter timescales. This sinusoidal variation has an amplitude of $\sim$9 lt-sec and a period of $\sim$760 d. This is indicative of presence of a third body around the compact X-ray binary. The mass and orbital radius of the third body are estimated to lie in the range, 20.5--26.9 Jupiter mass and 750-860 lt-sec, respectively. If true, then it will be the most massive circumbinary planet and also the smallest period binary known to host a planet.

\end{abstract}


\section{Introduction}

Low Mass X-ray Binary (LMXB) systems consist of a compact object accreting from a low mass companion star. The orbit of the LMXBs is expected to evolve due to mass transfer and redistribution of the angular momentum arising from the interaction of the binary components. 

The orbital evolution of X-ray binaries can be measured by four different ways. When the compact object is a pulsating neutron star, the pulse arrival time delay over the binary period is used to determine the orbital parameters of the system and multiple measurements of the orbital epoch over a long period are used to determine the orbital evolution \citep{Levine00}. Orbital evolution in some black hole X-ray binaries (BHXBs), has been measured by constructing the radial velocity curve of the companion star from the doppler shifts of the spectral lines \citep{Hernandez14, Hernandez17}. In the eclipsing binaries, connecting the mid eclipse time can give information on the long term evolution of the orbital period and an accurate determination of orbital period derivatives \citep{Wolff09, Jain10, Falanga15, Islam16}. The orbital evolution can also be measured from the stable orbital modulation of light curves \citep{Chou01a, Singh02}.

The orbital period of X-ray binaries can increase (for example, X 2127+119: \citet{Homer98}; SAX J1808.4--3658: \citet{Jain07}; 4U 1822--37: \citet*{Jain10} 4U 1916--053: \citet{Hu08}) or decrease (for example, 4U 1820-30: \citet{Chou01a}; Her X-1: \citet{Paul04}, \citet{Staubert09};  A 0620-00 \& XTE J1118+480: \citet{Hernandez14}; AX J1745.6-2901: \citet{Ponti16}; Nova Muscae 1991 \citet{Hernandez17}) smoothly over several years of measurements. The orbital period can also undergo distinct epochs of sudden change, as observed in EXO 0748-676 \citep{Wolff09} and XTE J1710--281 \citep{Jain11}.

\src~ is one of the rare LMXBs, that show X-ray eclipses in their light curves \citep{Cominsky89}. It is a transient X-ray source which was discovered in 1976 \citep{Lewin76} from observations made with the SAS-3 X-ray observatory. Through several follow-up observations, an orbital period of $\sim$7.1 hr and an eclipse duration of $\sim$15 min were determined \citep{Lewin78, Cominsky84}. About 2 years after its discovery, the X-ray intensity declined and the source was not detectable for the subsequent more than 20 years \citep{Zand99}. During another outburst and a renewed activity in 1999, burst oscillations with a period of $\sim$1.8 ms were reported \citep{Wijnands01}, which could be the spin period of the neutron star. After being X-ray bright for about 2.5 years, the source went into quiescence near the beginning of 2001. 

Comparing the orbital period of \src~ measured from two eclipses during the first outburst and from four eclipses during the early part of the second outburst, \citet{Wachter00} reported an orbital period decay and determined an average decay timescale of 10$^{7}$ yr. But since this source was not detectable for a long time in between the two outbursts, there is no detailed record of the orbital period changes. Later, \citet{Oosterbroek01} determined two more mid-eclipse times of \src, using the \emph{Beppo}-SAX data during the second outburst. These measurements, along with the previous values, however, were not compatible with a simple orbital decay as was suggested earlier by \citet*{Wachter00}. All the available eclipse measurements at this stage (eight), indicated some complexity in the orbital solution of this source. 

Recently, \src~ went into another outburst \citep{Negoro15}, thus enabling a definitive study of its orbital evolution. In this work, we have determined mid-eclipse times using newer \emph{RXTE}-PCA and \emph{XMM-Newton} observations of this source, made during the second and the current outburst. 
\section{Observations and Analysis}

The \emph{RXTE}-PCA consists of an array of five collimated proportional counter units with a total photon collection area of 6500 cm$^{2}$ \citep{Jahoda96}. We have analyzed 24 archived observations of \src~ made with the \emph{RXTE} observatory. The observation log is given in Table~1. The PCA data collected in the event mode was used to generate the light curves, using the \textsc{ftool}-\textsc{seextrct} from the astronomy software package \textsc{heasoft}-ver 6.10. The analysis was done in the energy band 2$-$20 keV. The background was estimated using the \textsc{ftool}-\textsc{pcabackest}. Faint source model was taken from the \emph{RXTE} website. Thereafter, barycentric corrections were applied to all X-ray timings. 

The \emph{XMM}-Newton Observatory \citep{Jansen01} carries three X-ray mirrors and three focal plane instruments, each with a field of view of about 30\arcmin $\times$ 30\arcmin. Complete X-ray eclipses of \src~ have been observed during two \emph{XMM}-Newton observations. We have analyzed both of these archived observations. The first of these observations was made during the second outburst. It lasted for $\sim$31.5 ks and covered two complete eclipses. Another $\sim$42.9 ks long observation during the current outburst covered one complete eclipse. Observation details are summarized in Table~1.  We have analyzed the 0.2-10 keV EPIC-PN data, using the XMM Science Analysis System (SAS version 8.0.0). Source counts were extracted from a circular region of radius 40\arcsec centered on the position of the target. Background events were extracted from a similar source-free circular region. Background subtracted light curves were barycenter corrected using the SAS tool \textsc{barycen}. Spectroscopic results from the February 2001 observation have been reported earlier by \citet{Sidoli01}. They have reported the presence of two eclipses. However, they did not report the mid-eclipse times for the purpose of orbital evolution measurement. 

\subsection{Eclipse Timing} 
\begin{figure}
\centering
\includegraphics[height=2.6in, width=1.3in, angle=-90]{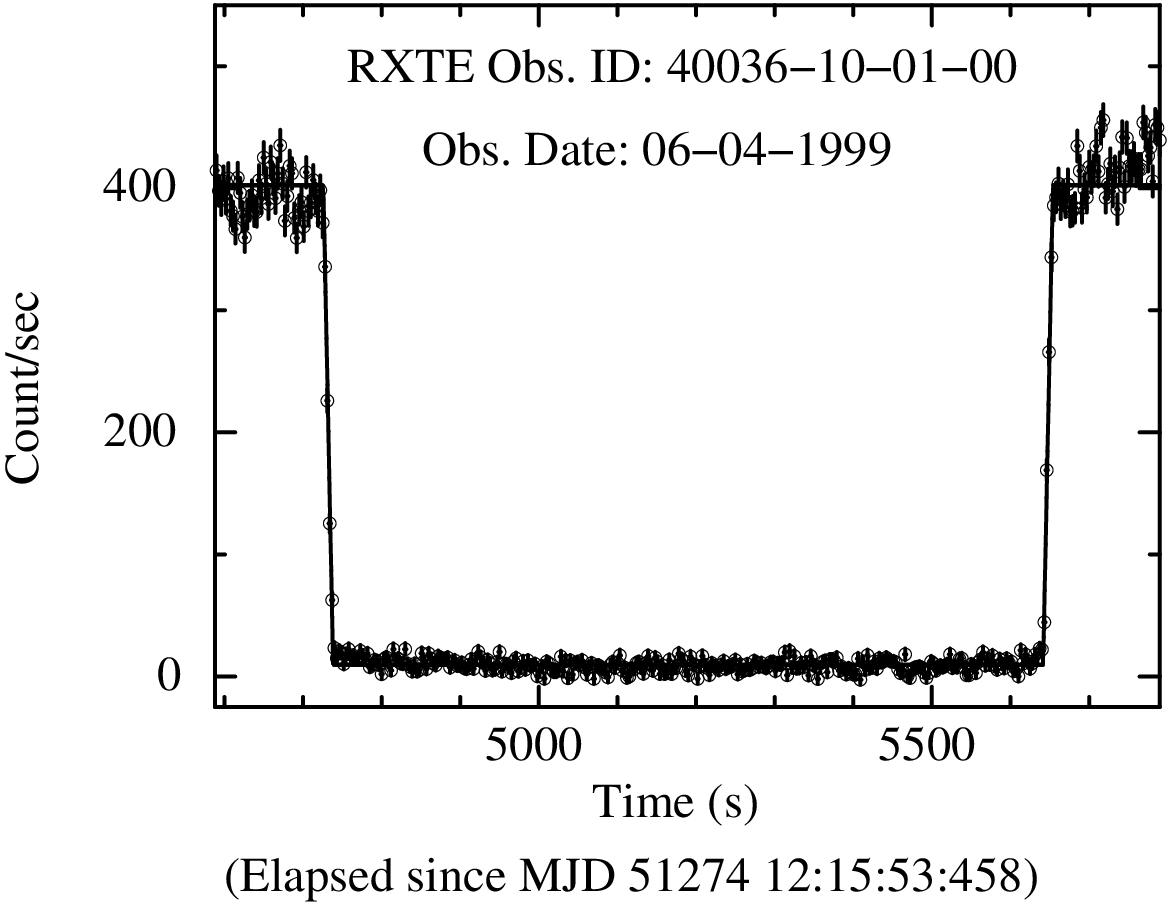}
\vspace{0.5 cm}

\includegraphics[height=2.6in, width=1.3in, angle=-90]{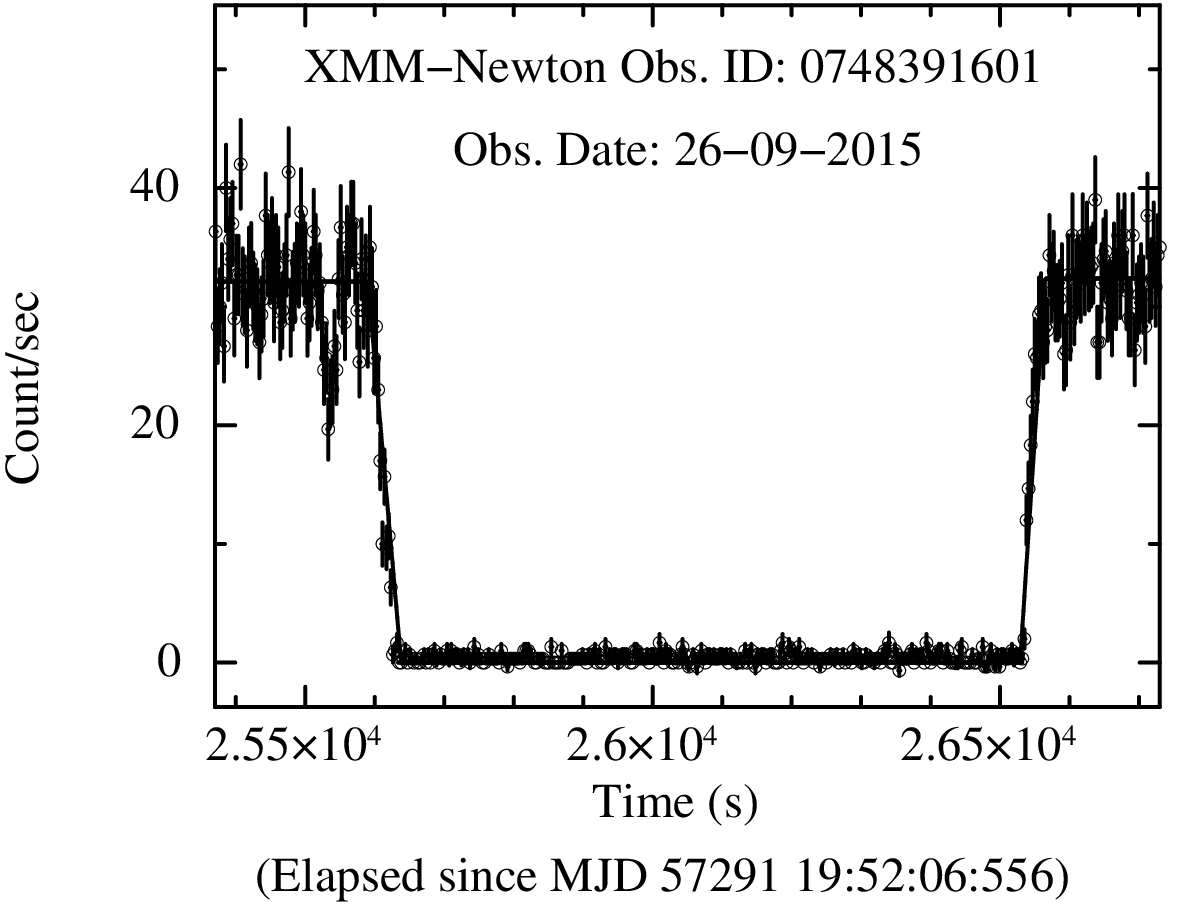}
\caption{Sample of background subtracted light curves of \src~ obtained from \emph{RXTE}-PCA and \emph{XMM}-Newton observations. The solid line in both the panels represents the best fit model as described in the text.}
\label{fig:sample}
\end{figure}
Figure~\ref{fig:sample} shows two sample background subtracted light curves of \src, binned with 3 seconds and including an eclipse lasting for $\sim$900 s. The mid eclipse times were determined by modeling each ingress and egress transition with a $``$step and ramp" model, which has been successfully employed in case of other eclipsing binaries \citep{Wolff09, Jain11}. The variable parameters of the model are, the pre-ingress, eclipse, and post-egress count rates; the ingress and egress duration, the eclipse duration and the mid-eclipse time. Considering all the components to be freely variable, we first fitted the seven-parameter model to the light curves covering the eclipse and $\sim$150 s before and after the eclipse (similar to \citet{Wolff09}). For \emph{RXTE}-PCA, it was found that the value of pre-ingress and the post egress count rates were similar and the eclipse ingress and egress duration were also similar within errors. The parameter space for \emph{RXTE}-PCA, was thus reduced to five.

From all the observations of \emph{RXTE} and \emph{XMM}-Newton, we have determined 27 mid-eclipse time measurements. The mid-eclipse times and the corresponding 1$\sigma$ statistical errors are given in Table~1. The durations of ingress and egress have been mentioned separately for the \emph{XMM}-Newton observations. The orbit numbers are in accordance with the eclipse time measurements given in \citet*{Wachter00}. As compared to other observations, the mid eclipse times determined from the \emph{RXTE} observations have smaller error, except in two observations where count rate was relatively low. 

Short timescale intrinsic variability in the intensity of the LMXB can modify the ingress/egress and thus effect the mid-eclipse time measurements. Therefore, in order to estimate the additional random error in the mid-eclipse time measurements due to variability in the source, we simulated eclipses (with similar parameters as given in Table~1) at several positions in an \emph{RXTE}-PCA light curve. Differences of the values of the mid-eclipse time that were used for simulation and those measured from the simulated data, were found to be about 1.3 seconds. Therefore, for further analysis, we have considered this value as an additional random error due to the intrinsic intensity variation of the LMXB, and have quadratically added it to the statistical error of each mid-eclipse time measurement.
\begin{table*}
\renewcommand{\thefootnote}{\alph{footnote}}
\centering
\begin{minipage}{200mm}
\caption{Measurements of the mid-eclipse times of \src.}
\begin{tabular}{@{}lclclcc@{}}
\hline
\hline
Observation 	& Instrument/		&	Observation 	&	Orbital	&	Mid eclipse time 	&	Duration of 		&	Duration of \\
Date	 	& Mission		&	ID		&	Cycle	&	MJD (d)\footnote{Number in brackets give the 1$\sigma$ statistical error.},\footnote{An independent random error due to intrinsic source variability has been added quadratically for further analysis.}&Eclipse (sec)\large{$^{a}$}	&Ingress/Egress (sec)\large{$^{a}$}\\
\hline
1976-10-07	&	SAS-3		&	--		&0	& 43058.72595 (15)\footnote{These numbers are taken from \citet{Cominsky89}}	&	--	&	--	\\ 
1978-03-07	&	HEAO-1		&	--		&1740	& 43574.64413 (15)\Large{$^{c}$}		&	--			&	--		\\ 
1999-04-05	&	RXTE		&	40414-01-01-00	&27707	& 51273.9780792 (15)\footnote{These numbers are taken from \citet*{Wachter00}}	&	--	&	--	\\
1999-04-06	&	RXTE		&	40036-10-01-00	&27709	& 51274.5711027 (37)  			&	903.10 (15)		&	12.78 (12)	\\
1999-04-09	&	RXTE		&	40414-01-02-00	&27720	& 51277.8326259 (37)\Large{$^{d}$}	&	--			&	--		\\
1999-04-10	&	RXTE		&	40050-04-02-00	&27722 	& 51278.4256585 (41)  			&	909.69 (15)		&	10.65 (5)	\\
1999-04-13	&	RXTE		&	40414-01-03-00	&27733	& 51281.6871743 (37)\Large{$^{d}$}	&	--			&	--		\\
1999-04-15	&	RXTE		&	40414-01-04-00	&27740	& 51283.7627259 (32)\Large{$^{d}$}	&	--			&	--		\\
1999-04-17	&	RXTE		&	40050-04-07-00	&27747 	& 51285.8382399 (21)  			&	899.24 (15)		&    	9.97 (20)	\\
1999-04-26	&	RXTE		&	40050-04-15-00	&27778 	& 51295.0298420 (58)  			&	905.50 (15)  		&   	9.788 (3)	\\
1999-04-29	&	RXTE		&	40050-04-16-00	&27787 	& 51297.6984644 (58)  			&	908.24 (15) 		&   	10.51 (4)	\\
1999-06-05	&	RXTE		&	40414-01-05-00	&27911  &51334.4649765 (23)  			&	898.51 (25)  		&   	8.55  (50)	\\
1999-06-06	&	RXTE		&	40414-01-06-00	&27915  &51335.6509729 (35) 			&	894.23 (20)  		&  	16.72  (56)	\\
1999-06-08	&	RXTE		&	40414-01-07-00	&27920  &51337.1334964 (42)  			&	898.72 (15)  		&   	10.69  (34)	\\
1999-08-03	&	RXTE		&	40414-01-09-00	&28110  &51393.4693583 (116)  			&	902.10 (80)  		&   	11.09  (32)	\\
1999-08-06	&	RXTE		&	40414-01-10-00	&28119  &51396.1378573 (41)  			&	898.76 (65)   		&  	12.42  (45)	\\
1999-08-07	&	RXTE		&	40414-01-11-00	&28123  &51397.3238583 (122) 	 		&	895.85 (70)   		&  	11.26  (38)	\\
1999-10-15	&	RXTE		&	40414-01-12-00	&28355  &51466.1129515 (23)  			&	904.77 (25)   		&  	9.78   (46)	\\
1999-10-16	&	RXTE		&	40414-01-14-00	&28359  &51467.2989903 (23)  			&	901.70 (15)   		&  	13.49  (31)	\\
1999-10-18	&	RXTE		&	40414-01-13-00	&28368  &51469.9675212 (23)  			&	898.99 (25)  		&   	16.04  (42)	\\
2000-01-14	&	RXTE		&	40414-01-15-00	&28663  &51557.4363538 (29)  			&	904.40 (20)   		&  	12.32  (31)	\\
2000-01-18	&	RXTE		&	40414-01-16-00	&28676  &51561.2909286 (12)  			&	909.02 (15)   		&  	9.48   (32)	\\
2000-01-19	&	RXTE		&	40414-01-17-00	&28680  &51562.4769833 (23)  			&	902.36 (35)   		&  	14.67  (53)	\\
2000-05-13	&	RXTE		&	50410-01-01-00	&29069  &51677.8173235 (23)  			&	901.50 (25)  		&   	16.26  (36)	\\
2000-05-17	&	RXTE		&	50410-01-02-00	&29082  &51681.6719031 (17)  			&	906.41 (15)   		&  	12.18  (23)	\\
2000-05-18	&	RXTE		&	50410-01-03-00	&29086  &51682.8579030 (58)  			&	902.26 (20)   		&  	13.59  (64)	\\
2000-08-08	&	RXTE		&	50410-01-06-00	&29363  &51764.9896893 (58)   			&	905.13 (20)   		&  	11.68  (48) 	\\
2000-08-12	&	Beppo-SAX	&	--		&29376 	&51768.844257 (16)\footnote{These numbers are taken from \citet{Oosterbroek01}} 	&	--&--		\\
2000-08-13	&	Beppo-SAX	&	--		&29378	&51769.437259 (15)\Large{$^{e}$}	&	--			&	--		\\
2000-10-18	&	RXTE		&	50410-01-07-00	&29600  &51835.2612753 (58)  			&	909.21 (15)    		& 	11.15 (44)  	\\
2000-10-19	&	RXTE		&	50410-01-08-00	&29604  &51836.4472795 (51)    			&	907.53 (10)  		&   	10.34 (15) 	\\
2000-10-20	&	RXTE		&	50410-01-09-00	&29606  &51837.0402811 (58)    			&	906.15 (15)  		&   	11.34  (35)  	\\
2001-02-20	&	XMM-Newton	&	0008620701	&30022	&51960.386091 (23)			&	903.13	(50)   		&  	15 (2)/ 15.5 (5) \\
2001-02-20	&	XMM-Newton	&	0008620701	&30023	&51960.682631 (28)			&	902.75	(84)  		&   	19 (1)/ 19.2 (1) \\
2015-09-26	&	XMM-Newton	&	0748391601	&48004	&57292.129569 (32)			&	900.95	(40)		& 	47 (3)/ 22 (4) 	\\
\hline
\end{tabular}
\end{minipage}
\end{table*}
\subsection{Results}
Since its discovery, \src~ has undergone three outbursts. The first outburst lasted from 1976 to 1978, the second phase between 1999 - 2001; and the current phase of enhanced emission started around August 2015. Only two mid eclipse time measurements have been reported during the first active period. And during the current active phase of \src, so far we have only one measurement of the mid eclipse time. In contrast, from the second outburst, we have a total of 32 mid-eclipse time measurements with \emph{RXTE}-PCA, \emph{Beppo}-SAX and \emph{XMM}-Newton.

\begin{figure}
\centering
\includegraphics[height=2.3in, width=1.5in, angle=-90]{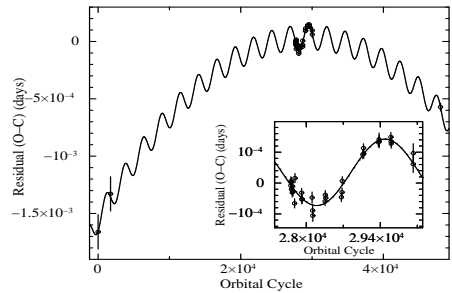}
\caption{The mid-eclipse times of \src~ and the best fit model are shown here after subtracting a linear fit to the data. The \textit{inset} figure shows enlarged view of the residuals during the second outburst.}
\label{fig:residual}
\end{figure}

We fitted a linear model to all the 35 mid eclipse time measurements. The best-fitting linear component was subtracted from the ephemeris history and the O-C residual curve is plotted in Figure~\ref{fig:residual}. This curve hints at an orbital period decay in the system. It is also evident that over and above a secular orbital period decay, this source shows a periodic residual at a shorter timescale. The pattern of the residual can not be fitted with a higher order polynomial. 

We fitted a model consisting of a quadratic and a sinusoidal function to the residual curve (Equation~\ref{eqn:eqn1}). 
\begin{eqnarray}
T_{n}=T_{0}+nP_{orb}+\frac{1}{2}n^{2}P_{orb}\dot{P}_{orb}+A_{sin}\sin\left(\frac{2\pi(n-n_o)}{P_{sin}}\right)
\label{eqn:eqn1}
\end{eqnarray}
In this equation, P$_{orb}$ is the orbital period at epoch T$_{0}$. The parameters, $A_{sin}$, $P_{sin}$ and $n_{0}$ are the amplitude, period and phase of the sinusoidal function, respectively. The best fit parameters are given in Table~2 and the best fit model is shown in Figure~\ref{fig:residual}, after subtracting the best fit linear model. The timescale for evolution of the orbital period ($\tau = P_{orb}$/--$\dot{P}_{orb}$) is 6.5(7) $\times$ 10$^{7}$ yr. It is larger than an earlier estimate using fewer mid-eclipse times \citep*{Wachter00} by a factor of $\sim$6. The sinusoidal variation in the O-C curve could be due to light travel time delay for motion of the X-ray binary in the presence of a third body. 
\begin{table}
\centering
\caption{Orbital ephemerides of \src}
\begin{tabular}{l l }
\hline
\hline
Parameter					&	Best fit value$^{a}$\\
\hline
T$_{0}$ (MJD)					&	43058.72606 (23)\\
P$_{orb}$ (d)					&	0.296504619 (11)\\
$\dot{P}_{orb}$					&	--1.25 (13)$\times$10$^{-11}$\\
$A_{sin}$ (lt-sec)				&	9.20 (75)	\\
P$_{sin}$ (d)					&	764 (37)	\\
n$_{0}$					&	28846 (28)	\\
$\chi^{2}$ (d.o.f)				&	46.65 (29)\\
\hline
\end{tabular}

$^{a}$ The numbers in bracket indicate the 1$\sigma$ errors
\end{table}
\section{Discussion}
We have determined 27 new mid-eclipse times of the X-ray binary \src~ using data from \emph{RXTE} and \emph{XMM}-Newton observatories. These measurements have been used to determine the orbital evolution in this system.

Orbital evolution of LMXBs is complex and is known to display different trends. The orbital separation is known to increase in most of the LMXBs, at timescales which are shorter than that predicted by a conservative mass transfer or by gravitational wave radiation \citep{Homer98, Hartman09}. A decreasing orbital period has been observed in a few LMXBs and some short period BHXBs. But the orbital decay in these systems is also unusual and is much faster than that predicted by conventional methods of gravitational wave radiation, magnetic braking and mass loss from the system \citep{Hernandez14, Ponti16}. Interaction with a third body could be responsible for a large orbital decay observed in two LMXBs \citep{Peuten14, Iaria15}. Orbital period glitches have been observed in a couple of LMXBs, and are attributed to be due to magnetic cycling of the secondary star \citep{Wolff09, Jain11}. Even though most of the LMXBs show a reasonably good quadratic fit to the mid-eclipse time records, there are signatures of deviation from a constant orbital period derivative present on longer timescales \citep{Iaria15, Chou16, Patruno16}.

The mid-eclipse time history of \src~ seems to be quite unusual. The combined data spanning three outbursts and covering $\sim$40 years of timeline, indicates an orbital decay.  In addition, a large number of measurements in a two year period during the second outburst show sinusoidal variation in the eclipse timing residual, perhaps indicating the presence of a third body around this source. 

We can consider the X-ray binary (the inner binary) to be a point mass in an approximate binary motion with this third body. In that case, the sinusoidal residual is due to orbital motion of the inner binary around the center of mass of the whole system. This is similar to the pulse arrival time delay of a binary X-ray pulsar, except that instead of a periodic pulse, we have a periodic eclipse.

For a 1.4 $M_{\odot}$ neutron star, the mass of the companion star lies between 0.25--0.9 $M_{\odot}$ \citep{Cominsky84}. Assuming the radius of the orbit of the X-ray binary (having an estimated total mass in the range 1.65--2.3 $M_{\odot}$) around the center of mass of the system to be same as amplitude of the sinusoidal residual, the third body (assumed to be co-planar with the inner binary) should have a mass range of 0.0195--0.0257 $M_{\odot}$ (i.e., 20.5--26.9 Jupiter mass) and an estimated range for orbital radius between 750--860 lt-sec. 

\begin{figure}
\centering
\includegraphics[height=2.5in, width=1.5in, angle=-90]{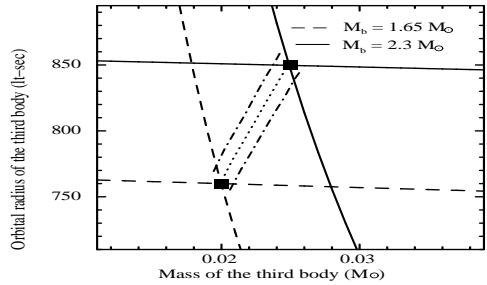}
\caption{Graphical representation of finding the mass and orbital radius of the third body. The square shaped markers indicate the mass and orbital radius of the third body for a binary mass of 1.65$M_{\odot}$ (dashed line) and 2.3$M_{\odot}$ (solid line), respectively.}
\label{fig:simulation}
\end{figure}

The two extremes of this estimation are graphically shown in Figure~\ref{fig:simulation}. Taking M$_{b}$ and R$_{b}$ as the mass and radius of orbit of the binary; and M$_{tb}$ and R$_{tb}$ as the mass and orbital radius of the third object, we have drawn curves for the expressions below for two extremes of the mass of the inner binary, i.e., 1.65 $M_{\odot}$ (dashed line) and 2.3 $M_{\odot}$ (solid line), respectively. (Here, G is the gravitational constant.)
\begin{eqnarray}
R_{tb} = \frac {M_{b} R_{b}} {M_{tb}};  
  R_{tb} = \left(\frac {G P_{sin}^2} {4 \pi^2}\right)^{1/3} \frac {M_{b}} {(M_{b}+M_{tb})^{2/3}}
\label{eqn:one}
\end{eqnarray}

In Figure~\ref{fig:simulation}, the square markers indicate these estimated parameters for two extremes of the mass of the inner binary. Depending on the mass of the inner binary, the true mass and orbital radius of the third body will lie on the dotted line connecting these square markers. The additional errors in these two parameters due to the uncertainty in period and amplitude of the sinusoidal component have been estimated by a Monte Carlo simulation. The additional 1$\sigma$ uncertainty is represented by two dash-dotted lines parallel to the diagonal (dotted) line.

The mass and orbital radius of the third body have been estimated assuming a circular and co-planar orbit. There is no evidence of circularity except that the residual shown in Figure 2 is consistent with being sinusoidal. Co-planarity is a reasonable assumption as the circumbinary planets discovered with Kepler are nearly co-planar \citep{Welsh14}

Though rarer compared to planets around single stars, about 20 circumbinary planets are known among $\sim$1000 eclipsing binaries observed with Kepler \citep{Welsh14}. If true, the third body in the present system is the most massive circumbinary planet; exceeding Kepler-1647b by a factor of about 15 \citep{Kostov16}. Simulations of planet formation and migration around binary stars show the most stable planets to be in the sub-Saturn mass range while more than Jupiter mass planets, if present, are likely to have large orbits around the binary \citep{Pierens08}.

The binary period of MXB 1658-298 is also much shorter compared to the orbital period of all the binary stellar systems around which planets have been found, the shortest binary period being 7.4 days in Kepler-47 \citep{Orosz12}. The lack of planets around short period binaries is believed to be related to the process of angular momentum loss that brings the two stars closer in the process of binary evolution \citep{Welsh14}.

The system MXB 1658-298 gives important new input for the range of stellar configurations for which circumbinary planets may form and survive migration over several stages of binary evolution. In particular, the binary system being a low mass X-ray binary and having an age of several billion years is an important input for the study of planet formation and migration around binary stellar systems.
\section*{Acknowledgments}
This research has made use of data obtained from the High Energy Astrophysics Science Archive Research Center, provided by NASA's Goddard Space Flight Center. The authors thank S. Sridhar for insightful discussions.

\label{lastpage}

\end{document}